\documentclass[twocolumn,trackchanges]{aastex631}

\newcommand{\msun}{$\mathrm M_{\odot}$}
\newcommand{\ngc}{NGC\,}
\newcommand{\hst}{\textit{HST}}

\newcommand{\afe}{\mathrm [\alpha/Fe]}

\accepted{November 26, 2024}

\submitjournal{ApJ}

\begin{document}
\title{Exploring Simple-Population and Multiple-Population Globular Clusters in the Outer Galactic Halo using the Hubble Space Telescope}

\correspondingauthor{E. Lagioia}
\email{elagioia@ynu.edu.cn}

\author[0000-0003-1713-0082]{E.~P. Lagioia}
\affiliation{South-Western Institute for Astronomy Research Yunnan University, Kunming, 650500, P.R. China} 
\author[0000-0001-7506-930X]{A.~P. Milone}
\affiliation{Dipartimento di Fisica e Astronomia ``Galileo Galilei'', Universit\`{a} di Padova, Vicolo dell'Osservatorio 3, 35122, Padova, Italy}
\affiliation{Istituto Nazionale di Astrofisica - Osservatorio Astronomico di Padova, Vicolo dell'Osservatorio, 5, 35122, Padova, Italy}
\author[0000-0001-7506-930X]{M.~V. Legnardi}
\affiliation{Dipartimento di Fisica e Astronomia ``Galileo Galilei'', Universit\`{a} di Padova, Vicolo dell'Osservatorio 3, 35122, Padova, Italy}
\author[0000-0002-7690-7683]{G. Cordoni}
\affiliation{Research School of Astronomy and Astrophysics, Australian National University, Canberra, ACT 2611, Australia}
\author[0000-0001-8415-8531]{E. Dondoglio}
\affiliation{Istituto Nazionale di Astrofisica - Osservatorio Astronomico di Padova, Vicolo dell'Osservatorio, 5, 35122, Padova, Italy}
\author[0000-0002-7093-7355]{A. Renzini}
\affiliation{Istituto Nazionale di Astrofisica - Osservatorio Astronomico di Padova, Vicolo dell'Osservatorio, 5, 35122, Padova, Italy}
\author[0000-0002-1128-098X]{M. Tailo}
\affiliation{Istituto Nazionale di Astrofisica - Osservatorio Astronomico di Padova, Vicolo dell'Osservatorio, 5, 35122, Padova, Italy}
\author[0000-0001-8538-2068]{T. Ziliotto}
\affiliation{Dipartimento di Fisica e Astronomia ``Galileo Galilei'', Universit\`{a} di Padova, Vicolo dell'Osservatorio 3, 35122, Padova, Italy}
\author[0000-0003-1757-6666]{M. Carlos}
\affiliation{Theoretical Astrophysics, Department of Physics and Astronomy, Uppsala University, Box 516, SE-751 20 Uppsala, Sweden}
\author[0000-0002-1562-7557]{S. Jang}
\affiliation{Center for Galaxy Evolution Research and Department of Astronomy, Yonsei University, Seoul 03722, Korea}
\author[0000-0002-1276-5487]{A.~F. Marino}
\affiliation{Istituto Nazionale di Astrofisica - Osservatorio Astronomico di Padova, Vicolo dell'Osservatorio, 5, 35122, Padova, Italy}
\author[0000-0001-5182-0330]{A. Mohandasan}
\affiliation{Dipartimento di Fisica e Astronomia ``Galileo Galilei'', Universit\`{a} di Padova, Vicolo dell'Osservatorio 3, 35122, Padova, Italy}
\author[0009-0007-8203-7753]{J. Qi}
\affiliation{South-Western Institute for Astronomy Research Yunnan University, Kunming, 650500, P.R. China} 
\author[0000-0002-6373-770X]{G. Rangwal}
\affiliation{South-Western Institute for Astronomy Research Yunnan University, Kunming, 650500, P.R. China}
\author{E. Bortolan}
\affiliation{Dipartimento di Fisica e Astronomia ``Galileo Galilei'', Universit\`{a} di Padova, Vicolo dell'Osservatorio 3, 35122, Padova, Italy}
\author{F. Muratore}
\affiliation{Dipartimento di Fisica e Astronomia ``Galileo Galilei'', Universit\`{a} di Padova, Vicolo dell'Osservatorio 3, 35122, Padova, Italy}
%


\begin{abstract}
The pseudo two-color diagram, known as {\it chromosome map} (ChM), is a valuable tool for identifying globular clusters (GCs) that consist of single or multiple stellar populations (MPs). Recent surveys of Galactic GCs using the ChM have provided stringent observational constraints on the formation of GCs and their stellar populations. However, these surveys have primarily focused on GCs at moderate distances from the Galactic center and composed of MPs.
In this paper, we present the first detailed study of the stellar composition of four GCs in the outer halo of the Milky Way: Arp\,2, Ruprecht\,106, Terzan\,7, and Terzan\,8. Our analysis is based on high-precision photometry obtained from images collected with the Hubble Space Telescope in the F275W, F336W, F438W, F606W, and F814W bands. We find that Ruprecht\,106 and Terzan\,7 are composed solely of a single stellar population, whereas Arp\,2 and Terzan\,8 host both first- and second-population stars. In these clusters, the second population comprises about half and one-third of the total number of GC stars, respectively.
The results from this paper and the literature suggest that the threshold in the initial GC mass, if present, should be smaller than approximately $10^5$\,\msun. The first-population stars of Arp 2 and Terzan 8, along with the stars of the simple-population GCs Ruprecht\,106 and Terzan\,7, exhibit intrinsic ${\rm F275W-F814W}$ color spreads, likely indicative of [Fe/H] variations of approximately 0.05 -- 0.30\,dex. This suggests that star-to-star metallicity variations are a common feature of star clusters, regardless of the presence of MPs. 

\end{abstract}



\section{Introduction}
In most Galactic globular clusters (GCs), stars exhibit variations in helium and light elements \cite[see][for reviews]{kraft94,bastian18,gratton12a,milone22}. In particular, the abundance of C and N, O and Na, and Mg and Al, show a negative pairwise correlation. Thus, N-rich stars are depleted in C and O, and enriched in He and Na. In fewer cases N-rich stars are also Al-rich and depleted in Mg \citep[e.g.][]{marino08,carretta09}.   

Old stars rich in C, O and Mg, are also commonly observed in the Galactic-halo field. It is for this reason that coeval GC stars with similar chemical composition are thought to be formed during the first star formation burst that followed the collapse of the giant molecular cloud from which their host cluster originated. This gives a rationale for calling them first population or 1P stars. 

On the other side, GC members enriched in He, N, Na, and Al are rarely detected among field stars. The chemical composition of these stars bear the footprints of hot proton-capture reactions triggered in the CNO, NeNa, and MgAl chains, which are activated at effective temperatures ($T_{\rm{eff}}$) $\gtrsim 75$\,MK \citep{denisenkov89b}. As a consequence, they are called second population or 2P stars, because short-lived 1P stars are regarded as the gas donors from which 2P stars formed. 2P stars can be therefore identified as a stellar generation subsequent to that of 1P stars \citep{cottrell81,dantona83a,yong05,denissenkov14}, or as 1P stars that incorporated material processed by massive stars through accretion or binary interaction \citep{demink09b,bastian13b,krause13,gieles18,renzini22}    
2P stars can constitute the main stellar component of massive GCs with MPs \citep{martell11a,milone20,dondoglio21,jang22} but they are infrequent in the field, so that halo stars with 2P-like chemistry are thought to be lost from GCs through tidal interaction with the Galaxy. 

Attempting to identify a comprehensive mechanism for the formation of MPs or, in simpler terms, discerning the nature of the 1P donors, encounters various observational constraints that are challenging to meet simultaneously \citep[for an extensive discussion on the various MP formation models and their implications see][]{renzini15a,renzini22}. Determining the relation between the properties of the multiple stellar populations (MPs) and the main parameters of the host cluster, such as the GC mass and the orbit, may provide constraints to the GC formation. To better refine our understanding of the formation scenarios, it is also crucial to ascertain whether MPs are a widespread trait among Galactic GCs or whether only clusters surpassing a specific mass threshold can sustain the formation of MPs.

Variations in abundance among stars contribute to a corresponding spread in colors in photometric diagrams obtained from specific filters \citep[e.g.][and references therein]{marino08,yong08,milone12a,lee15, mehta24, milone22}. As an example, the photometric diagram dubbed {\it chromosome map} (ChM) is an efficient and widely used tool to disentangle 1P and 2P stars along the main sequence (MS), red giant branch (RGB), and asymptotic giant branch (AGB) of star clusters. The ChM is a pseudo two-color diagram typically constructed with the pseudo-color $C_{F275W,F336W,F438W} = (m_{\rm F275W}-m_{\rm F336W})-(m_{\rm F336W}-m_{\rm F814W})$, mainly sensitive to internal variations in C, N and O; and the color $m_{\rm F275W}-m_{\rm F814W}$ which improves the identification of stars with varying helium content \citep{milone15a,marino19,lagioia18,lagioia21}. Various photometric surveys of Galactic GCs based on ChMs \citep{milone17,jang22} reveal the ubiquitous presence of MPs in all investigated clusters. These results support and reinforce similar conclusions drawn from spectroscopic analyses \citep{gratton04a,gratton12a, marino19}. 

In this context, the identification of simple-population (SP) GCs, namely clusters with chemically homogeneous stars, casts doubt on the universal occurrence of MPs, thus challenging the view of MPs as a phenomenon inherent to the proto-cluster environment. The best prototypes of SP GCs are Ruprecht\,106 and Terzan\,7. For the first cluster, the spectroscopic analyses by \citet{villanova13,francois14} and \citet{frelijj21} revealed homogeneous light-element abundance. Moreover, the analysis of the cluster HB morphology by \cite{milone14}, and the photometry in the F336W, F438W, and F814W bands by \citet{dotter18}, suggested that the color distribution of RGB stars was consistent with that of a single stellar population. Similarly, evidence based on ground-based $U, B, I$ photometry that the star cluster Terzan\,7 contains a single stellar population has been provided by \citet{lagioia19}.
  
Recently, we have been granted 62 orbits during the cycle 30 of the Hubble Space Telescope (\hst, GO 17075, PI: Lagioia) to investigate for the first time the ChMs of four outer-halo clusters. The targets include Ruprecht\,106 and Terzan\,7, the two aforementioned candidate SP GCs but never been studied through the ChM, which is mandatory for a conclusive assessment about either presence or absence of MPs. The other two targets are Arp\,2 and Terzan\,8.

The choice of our targets was based on the following criteria: galactocentric distance, metallicity, mass and origin. With respect to the first criterion, the galactocentric distance of all the targets exceeds 15\,kpc (\citeauthor{harris96a} \citeyear{harris96a}, 2010 update; \citeauthor{baumgardt21} \citeyear{baumgardt21}). 

As for the metallicity, the selected clusters span a wide range of iron content, ranging from the metal poor cluster Terzan\,8 \citep[$\rm{[Fe/H] \simeq -2.3}$\,dex;][]{mottini08,carretta14} to the sub-solar metallicity cluster Terzan\,7 \citep[$\rm{[Fe/H]=-0.62}$\,dex][]{tautvaisiene04}. We notice that the average metallicity values reported in the \citep[][2010 update]{harris96a} catalog span a similar interval, with [Fe/H]$=-2.18$\,dex for Terzan\,8 and [Fe/H]$=-0.32$\,dex for Terzan\,7. 

All the targets are relatively small clusters with present-day mass M $\leq 4 \cdot 10^4$\,\msun, except Terzan\,8, which is the most massive GC among them, with M $\sim 7.5 \times 10^4$\,\msun\ \citep[][2023 update]{baumgardt18a}. Baumgardt and collaborators have also provided an estimate of the initial mass of the target clusters~\footnote{In \citeyear{baumgardt18a}, \citeauthor{baumgardt18a} provided advanced estimates for the initial masses of the Galactic GCs. Their determinations face several uncertainties which depend on our limited understanding of the evolutionary history of the Milky Way, and therefore of its tidal field. Other uncertainties are related to the knowledge of the primordial mass segregation in clusters and their early chemo-dynamical evolution, in particular the amount of mass lost through interaction with the molecular clouds from which they originated. We notice that the aforementioned GC initial mass estimates are obtained by assuming a time-independent Galactic gravitational potential and no initial mass segregation.}, ${\rm M_{in}}$, that for three out of four clusters is smaller (albeit marginally in the case of Ruprecht\,106) than $\sim 1.5 \times 10^5$\,\msun, a threshold value proposed by \citet{milone20} for the onset of MPs. It is worth mentioning that according to studies based on the analysis of the CN and CH indices in low resolution spectra of relatively small and distant globulars, the initial mass threshold for the onset of MP might be lower than the value suggested by \citet{milone20} and approximately equal to $2 \times 10^4$\,\msun\ \citep{salinas15a,tang21,huang24}. 

Finally, concerning the last criterion, several studies indicate a probable ex-situ genesis of Arp\,2, Terzan\,7 and Terzan\,8, which are associated to the Sgr dSph \citep{bellazzini03,law10a,carretta14} and likely still bound to its main body \citep{bellazzini20a}. The origin of Ruprecht\,106 is doubtful: while previous studies rule out a link between this cluster and Sgr \citet[e.g.][]{pritzl05}, recent kinematic studies instead assess its association to the Sgr stream \citep{shirazi24}. Interestingly, \citet{sbordone07} showed that the chemical composition of Ruprecht\,106 is similar to that of the Sgr field stars. More recently \citet{villanova13,francois14} and \citet{lucertini23} have suggested an extra-galactic origin for Ruprecht\,106 based on the relative chemical abundance of $\afe$, r- and s-process elements in the analyzed stars, compared to those typical of GC and field stars of similar metallicity. 

The four target clusters represent, therefore, an ideal sample to study to which extent the environment affect the formation of MPs and to shed new light on the physical conditions driving the emergence of MPs. Table~\ref{tab:par} reports the main parameters of the four GCs.

The outline of the paper includes Section~\ref{sec:data}, where we describe the dataset and the data reduction techniques adopted in our analysis. Then, Section~\ref{sec:cmds} presents the UV-optical-NIR CMDs of the four targets clusters. In Section~\ref{sec:chms}, we present the first ChMs of our GCs, while in Section\,\ref{sec:1P} we investigate the metallicity of GC stars. Finally, the summary and the discussion are provided in Section\,\ref{sec:summary}. 

\begin{deluxetable*}{ccccccccc}
\tabletypesize{\small} 
\tablewidth{0pt}
\tablecaption{Main parameters of the four globular clusters analyzed in this work: cluster ID, galactocentric distance ($\rm R_{GC}$), iron content ([Fe/H]), reddening (E(B$-$V)), actual mass (M), initial mass (${\rm M_{in}}$), half-light radius (${\rm r_{hl}}$), escape velocity (${\rm v_{esc}}$), age. \label{tab:par}}
\tablehead{ \colhead{Cluster} & \colhead{${\rm R_{GC}}$} & \colhead{[Fe/H]} & \colhead{E(B$-$V)} & \colhead{M} & \colhead{${\rm M_{in}}$} & \colhead{${\rm r_{hl}}$} & \colhead{${\rm v_{esc}}$} & \colhead{Age} \\
 \colhead{} & \colhead{(kpc)} & \colhead{(dex)} & \colhead{(mag)} & \colhead{($10^4$\,\msun)} & \colhead{($10^5$\,\msun)} & \colhead{(arcmin)} & \colhead{(${\rm km\,s^{-1}}$)} & \colhead{(Gyr)}}  
\startdata
Arp\,2        & 21.39 & -1.75 & 0.10 & 3.87 & 0.912 & 1.70 & 4.5 & 13.00 \\   
Ruprecht\,106 & 18.04 & -1.68 & 0.20 & 3.40 & 1.259 & 1.26 & 5.6 & 11.50 \\
Terzan\,7     & 16.85 & -0.32 & 0.07 & 2.23 & 0.525 & 0.90 & 4.6 & 8.00 \\
Terzan\,8     & 20.43 & -2.16 & 0.12 & 7.59 & 7.586 & 1.89 & 6.0 & 13.50 \\
\enddata
\tablecomments{For our analysis, we adopted the average metallicity values from \citet[][2010 update]{harris96a}. The listed reddening values have been also taken from the same catalog. It is worth noting that \citet{mottini08} found ${\rm [Fe/H]=-1.83}$ for Arp\,2, while \citet{villanova13} obtained ${\rm [Fe/H]=-1.47}$ for Ruprecht\,106. Ages are taken from \citet{dotter10} except for Ruprecht\,106, for which the age comes from \citet{dotter18} . All other quantities listed in the table are derived from the 2023 updated version of the catalog of the Fundamental Parameters of Galactic Globular Clusters available at the following URL: \url{https://people.smp.uq.edu.au/HolgerBaumgardt/globular/}.}
\end{deluxetable*}
    
\section{Data analysis} \label{sec:data}
The observation dataset analyzed in this work for Arp\,2, Terzan\,7, and Terzan\,8 consists of images collected by \hst\ with the WFC3 camera in the bands F275W, F336W, and F438W of the UVIS channel during the cycle 30 (GO 17075; PI: Lagioia), and of ACS/WFC images in the bands F606W and F814W of ACS/WFC, already present on the MAST archive~\footnote{\url{https://archive.stsci.edu/}} before the execution of the GO-17075 program. For Ruprecht\,106, only F275W observations have been collected during the GO-17075, whereas the F336W and F438W images are available from the archive. The \hst\ data collected for the GO-17075 and analyzed in this paper can be found in MAST: \dataset[10.17909/23y6-1r51]{http://dx.doi.org/10.17909/23y6-1r51}. Details on the exposures used in this works are outlined in Table~\ref{tab:dataset}.

\begin{deluxetable*}{cccccc}
\tabletypesize{\scriptsize} 
\tablecaption{Dataset of \hst\ observations analyzed in this work. \label{tab:dataset}}
\tablehead{\colhead{Cluster} & \colhead{Date (DD/MM/YY)} & \colhead{Camera/Channel} & \colhead{Filter} & \colhead{N $\times$ exposure time (s)} & \colhead{Proposal ID (PI Last name)}}
\startdata
       & March 06--17 2023  & WFC3/UVIS & F275W & $25\times1325$                         & 17075 (Lagioia) \\
       & April 25--26 2023  & WFC3/UVIS & F275W & $2\times1325$                          & 17075 (Lagioia) \\
       & March 12--17 2023  & WFC3/UVIS & F336W & $11\times795$                          & 17075 (Lagioia) \\
Arp\,2 & March 12--17 2023  & WFC3/UVIS & F438W & $11\times348$                          & 17075 (Lagioia) \\
       & May 03 2016        & ACS/WFC   & F606W & $45+2 \times (554+555) + 4 \times 556$ & 14235 (Sohn) \\
       & April 22 2006      & ACS/WFC   & F606W & $40 + 5\times 345$                     & 10775 (Sarajedini) \\
       & April 22 2006      & ACS/WFC   & F814W & $40 + 5\times 345$                     & 10775 (Sarajedini) \\
\hline
              & June 29--30 2023   & WFC3/UVIS & F275W & $2 \times (1401+1402)$                & 17075 (Lagioia) \\
              & July 07--12 2023   & WFC3/UVIS & F275W & $12 \times (1401+1402)$               & 17075 (Lagioia) \\
              & December 10 2016   & WFC3/UVIS & F336W & $2 \times 1100$                       & 14726 (Dotter) \\
              & March 29 2017      & WFC3/UVIS & F336W & $2 \times 1100$                       & 14726 (Dotter) \\
              & May 30--31 2017    & WFC3/UVIS & F336W & $2 \times 1100$                       & 14726 (Dotter) \\
              & September 29 2017  & WFC3/UVIS & F336W & $4 \times 1100$                       & 14726 (Dotter) \\
Ruprecht\,106 & December 10 2016   & WFC3/UVIS & F438W & $571$                                 & 14726 (Dotter) \\
              & March 29 2017      & WFC3/UVIS & F438W & $571$                                 & 14726 (Dotter) \\
              & May 31 2017        & WFC3/UVIS & F438W & $571$                                 & 14726 (Dotter) \\
              & September 21 2017  & WFC3/UVIS & F438W & $2 \times 571$                        & 14726 (Dotter) \\
              & July 04 2010       & ACS/WFC   & F606W & $55 + 4 \times 550$                   & 11586 (Dotter) \\
              & July 12 2016       & ACS/WFC   & F606W & $60 + 841 + 844 + 2 \times (842+845)$ & 14235 (Sohn) \\
              & July 04 2010       & ACS/WFC   & F814W & $60 + 3 \times 585 + 586$             & 11586 (Dotter) \\ 
\hline
          & March 08--16 2023 & WFC3/UVIS & F275W & $17 \times 1325$                         & 17075 (Lagioia) \\
          & March 14--15 2023 & WFC3/UVIS & F336W & $3 \times 795 + 837 +2 \times 838$       & 17075 (Lagioia) \\
Terzan\,7 & March 14 2023     & WFC3/UVIS & F438W & $3 \times 348$                           & 17075 (Lagioia) \\
          & June 03 2006      & ACS/WFC   & F606W & $40 + 5 \times 345$                      & 10775 (Sarajedini) \\
          & May 04 2016       & ACS/WFC   & F606W & $45 + 2 \times (554+555) + 4 \times 556$ & 14235 (Sohn) \\ 
          & June 03 2006      & ACS/WFC   & F814W & $40 + 5 \times 345$                      & 10775 (Sarajedini) \\
\hline
          & March 09--23 2023 & WFC3/UVIS & F275W & $9 \times 1325$                          & 17075 (Lagioia) \\
          & April 18--30 2023 & WFC3/UVIS & F275W & $15 \times 1325$                         & 17075 (Lagioia) \\
          & May 01 2023       & WFC3/UVIS & F275W & $1325$                                   & 17075 (Lagioia) \\
          & March 16 2023     & WFC3/UVIS & F336W & $795$                                    & 17075 (Lagioia) \\
          & April 24--30 2023 & WFC3/UVIS & F336W & $9 \times 795$                           & 17075 (Lagioia) \\
Terzan\,8 & May 01 2023       & WFC3/UVIS & F336W & $795$                                    & 17075 (Lagioia) \\
          & March 16 2023     & WFC3/UVIS & F438W & $348$                                    & 17075 (Lagioia) \\
          & April 24--30 2023 & WFC3/UVIS & F438W & $9 \times 348$                           & 17075 (Lagioia) \\
          & May 01 2023       & WFC3/UVIS & F438W & $348$                                    & 17075 (Lagioia) \\
          & June 03 2006      & ACS/WFC   & F606W & $40 + 5 \times 345$                      & 10775 (Sarajedini) \\
          & April 28 2016     & ACS/WFC   & F606W & $45 + 2\times (554+555) + 4 \times 556$  & 14325 (Sohn) \\
          & June 03 2006      & ACS/WFC   & F814W & $40 + 5 \times 345$                      & 10775 (Sarajedini) \\  
\enddata
\end{deluxetable*}

The data reduction has been performed with the software \textsc{img2xym} \citep{anderson06}. The software takes as input the {\it flc} images, which are \hst\ pipeline-calibrated individual exposures corrected for pixel-based charge transfer efficiency \citep{anderson10a}. For each analyzed image, a $5\times5$ array of perturbed point-spread functions (PSFs) is generated starting from library empirical PSF models, and improved with spatial-variation corrections obtained from unsaturated and isolated bright stars. Saturated stars are also identified and their magnitude is accurately determined considered flux bleed into adjacent pixels \citep{anderson08,gilliland10}.      

Instrumental magnitudes were calibrated to the VEGA-MAG system using the method of \cite{bedin05}, incorporating UVIS and WFC encircled energy distribution and photometric zero points from the STScI website. Geometric distortion corrections were applied to star positions using the solution presented by \citet{bellini11}, and the positions were transformed to the Gaia DR3 reference system \citep{gaiaDR3_23}.
Finally, we selected stars measured with high photometric accuracy, following the methodology outlined in \citet{milone09} which is based on the {\it q-fit}, a quality index provided by the software.

To determine the cluster membership for each target, we took advantage of the large temporal baseline of our observations and compute proper motions (PMs). The adopted method is based on the  displacement of stars observed at different epochs with respect to the motion of a sample of reference stars. Since cluster members have a negligible proper-motion dispersion 
compared to field stars, we referred the PMs to a sample of bright, unsaturated cluster stars that are well fitted by the PSF model. We refer to the papers by \citet{anderson06}, \citet{lagioia14a}, and \citet{tailo21} and references therein for details on the procedure.

As shown by \citet[][see their figure 1]{tailo21}, who derived the stellar PMs of stars in the fields of view studied in this paper by using the \hst\ data available at that time, the field stars and the cluster members are well separated in the PM diagrams and the residual field contamination is negligible. 

As a final step, we verified that the photometry of each cluster is not significantly affected by differential reddening. Hence, we corrected the photometry for the effects of spatial-dependent variation of the photometric zero point, which are due to small inaccuracies in the PSF models, by applying the procedure outlined in \citet{milone12c}.

Overall, the dataset that we collected for analyzing MPs in our four outer-halo GCs is similar to that adopted in the recent surveys of Galactic GCs \citep{piotto15, milone17}, which are based on similar data in F275W, F336W, F438W, F606W, and F814W of the WFC3/UVIS and ACS/WFC cameras of \hst. Since we have derived stellar photometry and astrometry using the same methods and computer programs adopted in previous surveys, we can directly compare our results with former analyses focused on MPs in GCs, taking advantage of a sample composed by 62 Galactic GCs. 

\section{The UV-optical-NIR CMDs} \label{sec:cmds}
\begin{figure*}
\centering
\includegraphics[clip, trim=0.7cm 5cm 1cm 3cm,width=\textwidth]{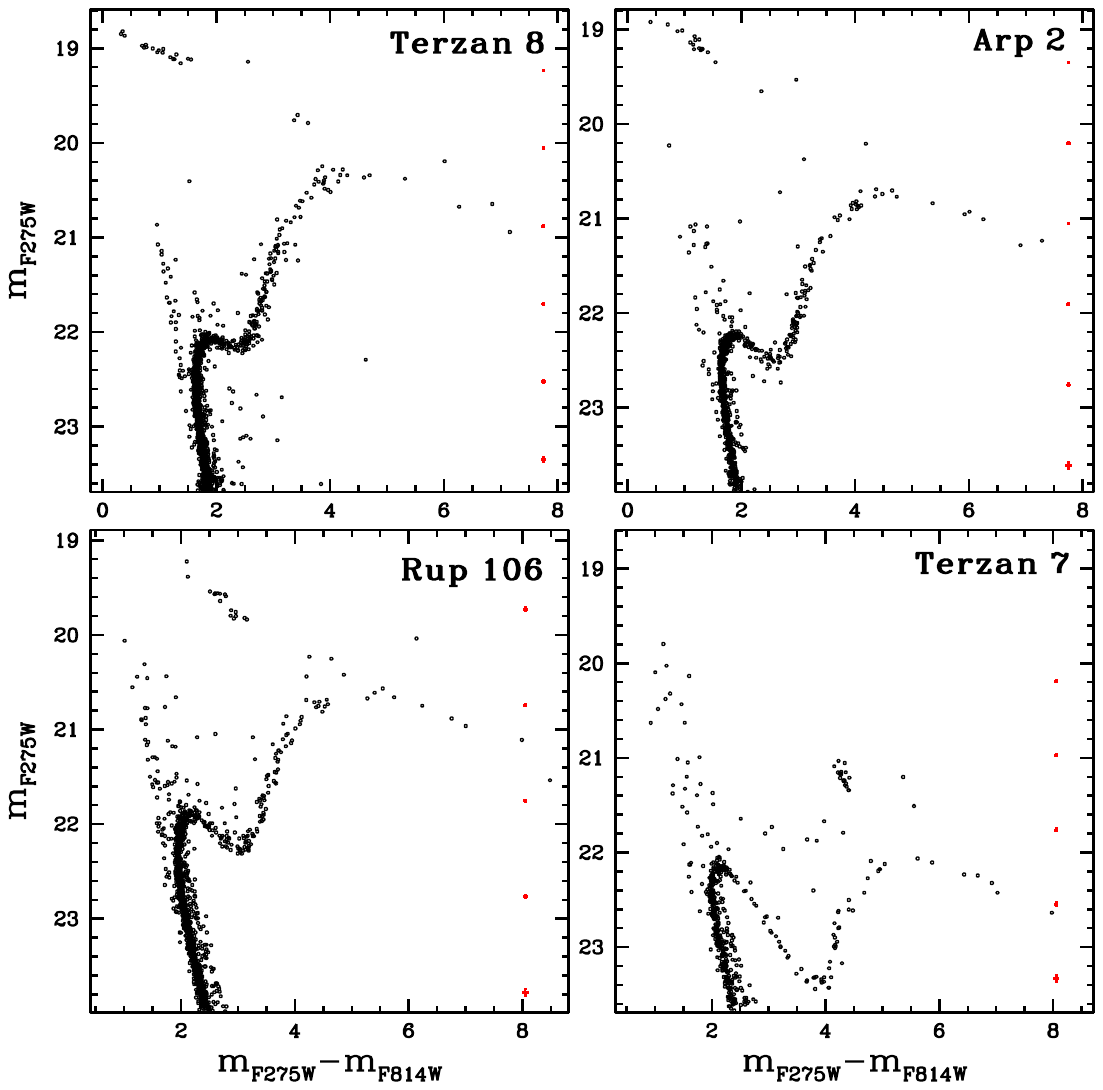}
\caption{$m_{\rm F275W}$ versus $m_{\rm F275W}-m_{\rm F814W}$ CMDs of the four clusters analyzed in this work, arranged by increasing metallicity from left to right and top to bottom. In each panel, the vertical and horizontal error bars on the right-hand side represent the typical magnitude and color errors at different F275W magnitudes. \label{fig:cmds_1}}
\end{figure*}

The CMDs built with the $m_{\rm F275W}-m_{\rm F814W}$ and $C_{\rm F275W,F336W,F438W}$ color indices are efficient tools to detect MPs. To this purpose, we plot in Figure~\ref{fig:cmds_1} the $m_{\rm F275W}$ versus $m_{\rm F275W}-m_{\rm F814W}$ CMDs of our four targets, from the most metal-poor (top-left) to the most metal-rich (bottom-right). In each panel, the error bars on the right side indicate the typical magnitude and photometric color error at the corresponding F275W magnitude.

We see that the faint region of each CMD is populated by MS stars reaching $\sim 1$ to 2\,mag below the Turn-Off (TO). We also observed conspicuous populations of blue stragglers and prominent sequences of MS-MS binary systems with large mass ratios, which appear on the red side of the MS in each cluster. The upper CMD region is populated by RGB, Horizontal-Branch (HB) stars and AGB stars. It is worth noticing the short color distribution of HB stars in all the target GCs. While theoretical prescriptions imply that young and metal-rich GCs display short HBs, the same phenomenon observed in intermediate-metallicity GCs might serve as an indirect indication that there is negligible to no helium enhancement among the cluster stars, as in the case of Ruprecht\,106 \citep{milone14}. Among the analyzed GC, the most metal-poor targets Arp\,2 and Terzan\,8 display slightly more extended HBs (${\mathbf \Delta(m_{\rm F275W}-m_{\rm F814W})} \sim 1.5$\,mag). 

\begin{figure*}
\centering
\includegraphics[clip,trim=0.7cm 5cm 1cm 3cm,width=\textwidth]{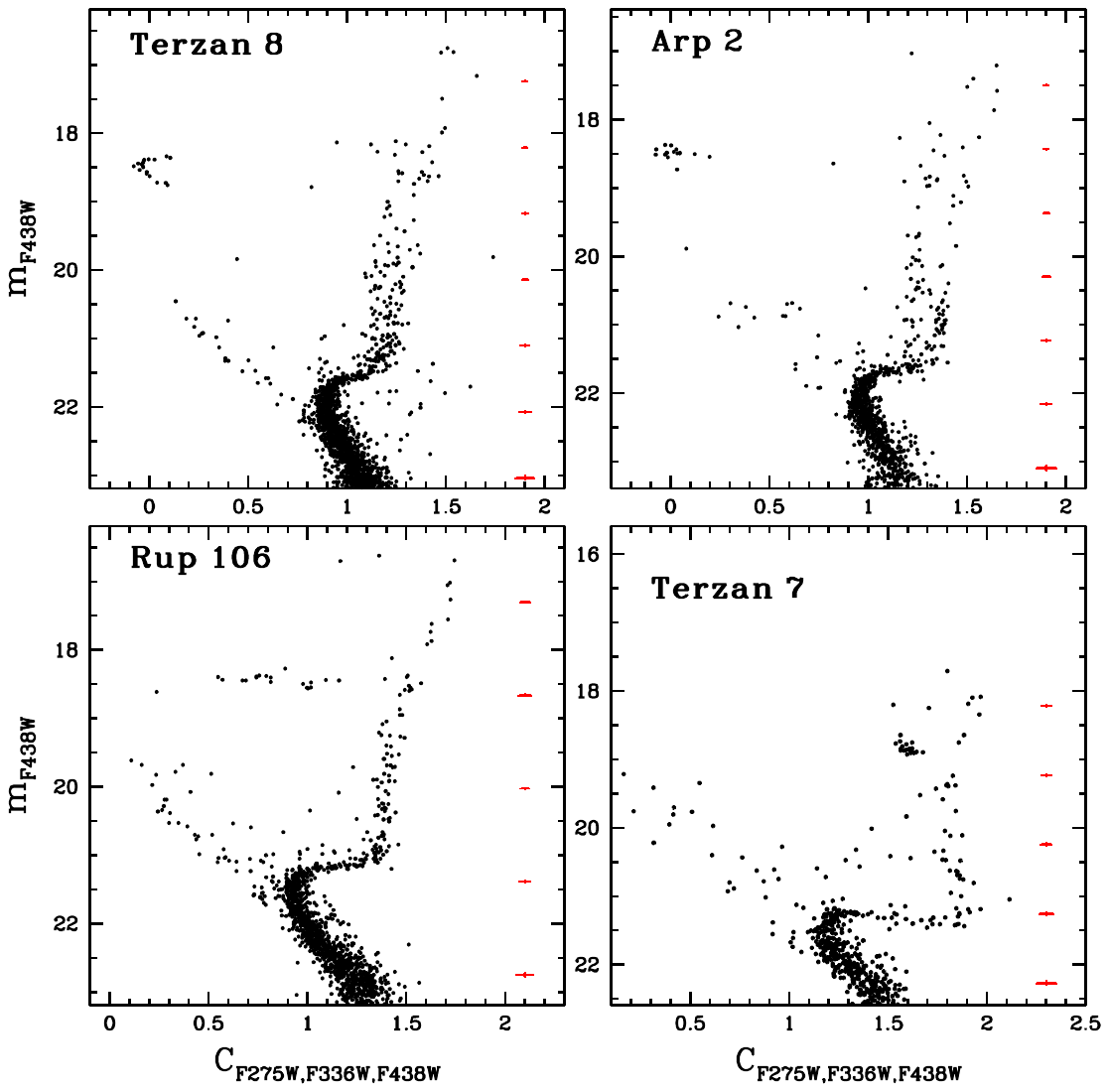}
\caption{$m_{\rm F438W}$ versus $C_{\rm F275W,F336W,F438W}$ pseudo-CMDs of the four target clusters, arranged as in Fig~\ref{fig:cmds_1}. In each panel, the error bars on the right side indicate the typical color and magnitude errors at different F438W magnitudes.\label{fig:cmds_2}}
\end{figure*}

Figure~\ref{fig:cmds_2} displays the $m_{\rm F438W}$ versus $C_{\rm F275W,F336W,F438W}$ pseudo-CMDs of our targets, arranged as in Fig.~\ref{fig:cmds_1}. The pseudo-CMDs of the two GCs in the top row show the largest color spread among the four targets. In particular, we observe a distinct split along the entire RGB extension in Arp\,2, whereas the RGB stars of Terzan\,8 with comparable magnitudes display a broad color spread, with indications of a split.

In both the cases, the $C_{\rm F275W,F336W,F438W}$ pseudo-color broadening of RGB stars is taken as evidence that these clusters harbor MPs. A spurious effect due to random errors in the pseudo-color is ruled out by the size of the color error bars. On the other side, the RGB spread visible in the CMDs of Ruprecht\,106 and Terzan\,7 is comparable with the error in color along the whole RGB extension. This suggests that Arp\,2 and Terzan\,8 harbor MPs with different light-element abundances, while Ruprecht\,106 and Terzan\,7 may not. 

\section{The {\it chromosome maps}} \label{sec:chms}
To further investigate the stellar populations along the RGB of the studied GCs, we constructed a pseudo color-color diagram called ChM. In this diagram, stars with different content of light elements such as helium, carbon, nitrogen, oxygen, and magnesium, occupy specific regions \citep{milone17,marino19a}. As a result, a ChM gives the possibility to classify large numbers of stars into various populations and with an accuracy competing with that provided by high-resolution spectroscopy \citep{marino19}.
In the subsequent section, we briefly describe the method used to derive the ChMs of the studied GCs, whereas in section~\ref{sub:chmMPs} we presents the analysis of the ChMs and results on the fraction of 1P and 2P stars in the target clusters.
 
 \subsection{Constructing the chromosome maps}\label{sub:chm}
A ChM displays along the abscissa and ordinate the indices $\Delta_{\rm F275W,F814W}$ and $\Delta{C_{\rm F275W,F336W,F438W}}$, respectively. Specific details on the procedure to derive the ChM have been widely discussed in various papers of our team \citep[e.g.][]{milone17,lagioia21}, to which we refer the interested reader. 

In a nutshell, to derive the ChM abscissa we take advantage of the $m_{\rm F814W}$ versus $m_{\rm F275W}-m_{\rm F814W}$ CMD, and of the $m_{\rm F814W}$ versus $C_{\rm F275W,F336W,F438W}$ pseudo-CMD to derive the ChM ordinate. In both these diagrams, we measure the (pseudo-) color displacement of every RGB star relative to the red boundary of the RGB sequence, and divide it by the broadening of the RGB sequence. The RGB broadening is defined as the (pseudo-) color difference between the red and the blue RGB boundaries at the same F814W star's magnitude. This result of this process, called verticalization, is a normalized color ratio that is finally multiplied by a factor called intrinsic RGB width.

To derive the intrinsic RGB width, we first measure the broadening of the RGB, or observed RGB width, at a reference F814W magnitude set by convention at $m^{\rm MSTO}-2$~\footnote{The F814W luminosity of the MSTO has been determined by using the naive estimator method \citep{silverman86}. First, we selected a magnitude interval in the $m_{F814W}$ versus $m_{F438W}-m_{F814W}$ CMD which includes MS and sub-giant branch (SGB) stars, and divided it into a given number of magnitude bins. Then, we computed the median color and magnitude of the stars in each bin. The whole procedure has been repeated $n$-times, with $n$ assuming a different value for each cluster ($n$ = 5, 6, 8), keeping the bin-width unchanged but shifting the starting point of the first bin by an amount equivalent to 1/$n^{th}$ of the bin-width. For each bin, the resulting median points have been linearly interpolated across the entire initial magnitude interval thus obtaining a raw fiducial line, which has been finally smoothed by boxcar averaging three adjacent points. Finally, the magnitude of the bluest point of the smoothed fiducial line has been taken as the cluster MS turn-off luminosity in F814W band. A similar approach has been adopted for the determination of the RGB boundaries. The blue and red RGB envelopes correspond, respectively, to the fiducial lines interpolating the $4^{\rm th}$ and $96^{\rm th}$ percentile of the RGB (pseudo-) color distribution in each bin, across the whole magnitude range covered by the cluster RGB members. Their relative distance of the distance at two F814W magnitudes above the MSTO correspond to the observed RGB width.}. Then, since the observed RGB width encompasses both the intrinsic RGB width and the observational error, it is necessary to estimate the contribution of the latter to the observed RGB width. To do that, we simulated the $m_{\rm F814W}$ versus $m_{\rm F275W}-m_{\rm F814W}$ and $m_{\rm F814W}$ versus $C_{\rm F275W,F336W,F438W}$ diagrams for a SP composed of 10,000 stars. The errors associated to each magnitude of the simulated stars are extracted from a Gaussian distribution with a dispersion corresponding to the median magnitude error of the RGB stars. Finally, the intrinsic RGB width in the color $m_{\rm F275W}-m_{\rm F814W}$ and in the pseudo-color $C_{\rm F275W,F336W,F438W}$, indicated respectively as $W_{\rm F275W,F814W}$ and $W_{C\,{\rm F275W,F336W,F438W}}$, have been obtained by subtracting in quadrature the RGB widths of the simulated CMD from observed RGB width ones. 

To evaluate the statistical error affecting our determination of $W_{\rm F275W,F814W}$ and $W_{C{\rm F275W,F336W,F438W}}$, we applied a bootstrapping test with replacement 1,000 times to the observed RGB colors and magnitudes. For each system configuration, we measured the intrinsic RGB width. We calculated the median value of the 1,000 RGB-width determinations, and the residuals from the median. We assumed that the $68^{\rm th}$ percentile of the distribution of the absolute values of the residuals corresponds to the uncertainty associated to the intrinsic RGB widths $W_{\rm F275W,F814W}$ and $W_{C{\rm F275W,F336W,F438W}}$. The values of the intrinsic width and the statistical error for the four targets GCs are reported in Table~\ref{tab:values}.    

\begin{figure}
\centering
\includegraphics[clip,trim=.5cm 5.0cm 1.7cm 3cm,width=\columnwidth]{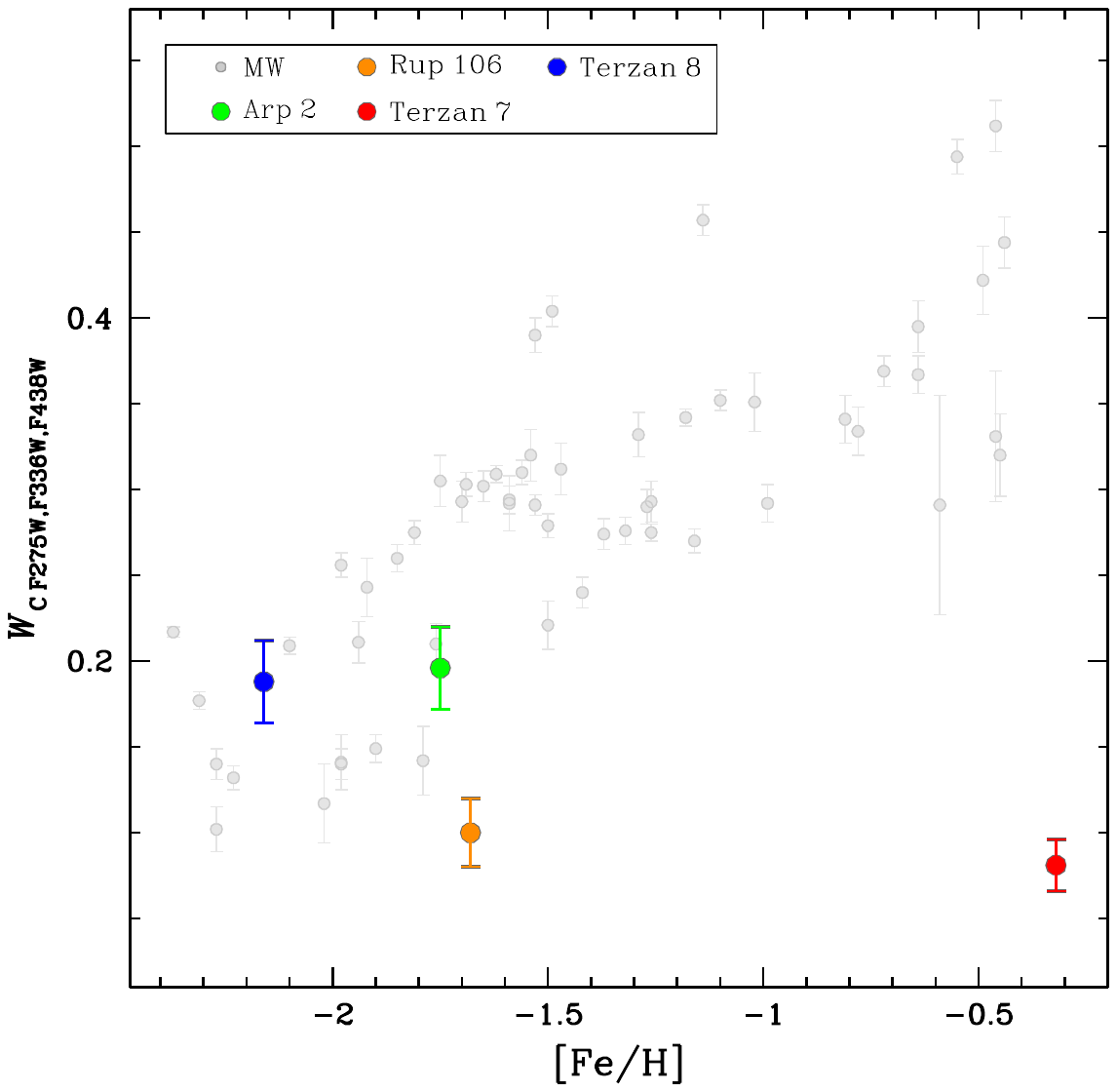}
\caption{RGB width in the pseudo-color $C_{\rm F275W,F336W,F438W}$ as function of iron abundance. The gray points represent the 58 MW clusters analyzed in \citet{milone18}. The four target clusters studied in this paper are represented as colored points, as reported in the inset. \label{fig_w}}
\end{figure}

The estimate of the RGB width of our target GCs allow us to contextualize these four clusters within the broader trend of absolute chemical variations observed in previously studied Galactic GCs \citep{milone17,lagioia19a}. To this purpose, Figure~\ref{fig_w} shows the intrinsic $W_{C\,{\rm F275W,F336W,F438W}}$ as a function of the average iron abundance for the 58 Galactic GCs studied by \citet[][gray dots]{milone18b} and the four outer-halo GCs studied in this paper (colored symbols). We find that Terzan\,8 and Arp\,2 follow the $W_{C\,{\rm F275W,F336W,F438W}}$ - [Fe/H] correlation of the bulk of studied GCs. On the other hand, Ruprecht\,106 and Terzan\,7 are remarkable exceptions, having RGB width smaller than those of GCs at similar metallicities. This observation supports the indication  from the color extension in the $m_{\rm F814W}$ versus ${\rm C_{F275W,F336W,F438W}}$ pseudo-CMD in Fig.~\ref{fig:cmds_2} that the two latter clusters may not host MPs. In this regard a definitive conclusion is provided in the next section, where the detailed stellar composition of the target clusters is explored through the ChM. 

\subsection{Multiple stellar populations in the chromosome map}\label{sub:chmMPs}
\begin{figure*}
\centering
\includegraphics[clip,trim=0.7cm 5.8cm 0.5cm 13cm,width=\textwidth]{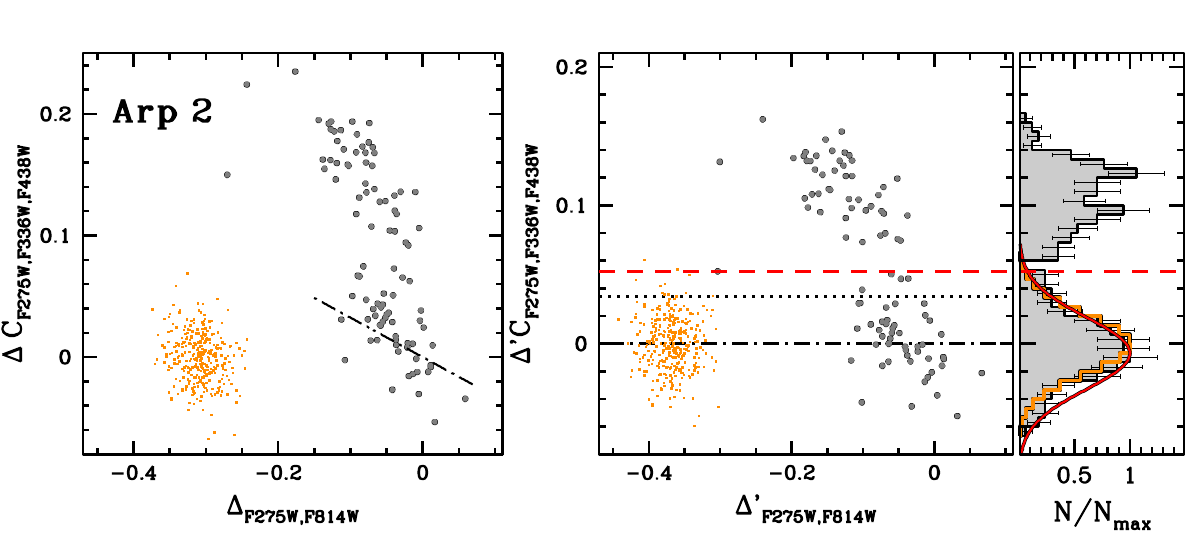}
\caption{Procedure for the selection of the candidate 1P and 2P stars of Arp\,2. \emph{Left panel}: $\Delta {C_{\rm F275W,F336W,F438W}}$ versus $\Delta_{\rm F275W,F814W}$ ChM of the cluster RGB stars (gray points). The orange dots represent the photometric error distribution of the ChM. The dot-dashed line marks the inclination of 1P stars with respect to the x-axis and has an inclination of $18^{\circ}$. \emph{Middle panel}: ChM and error distribution rotated counter-clockwise by an angle of $18^{\circ}$. In the rotated reference system $\Delta'{C_{\rm F275W,F336W,F438W}}$ versus $\Delta'{\rm F275W,F814W}$, the dot-dashed line is parallel to the x-axis. The dotted line marks the $95^{\rm th}$ percentile of the photometric error distribution along the y-axis. \emph{Right panel}: vertical histogram distribution of the ChM (gray-filled area) and photometric errors (orange histogram). The red curve represents the best-fit Gaussian to the region of the histogram delimited by the dotted line (see text for explanation). The red dashed line is the vertical coordinate corresponding to twice the dispersion of the best-fit Gaussian above its peak, and marks the separation between bona-fide 1P and 2P cluster stars. \label{fig:1P_2P_sel}}
\end{figure*}

As mentioned in the beginning of section~\ref{sec:chms} , the ChM allow us to tag the different stellar populations in a GC. As an example, in Figure~\ref{fig:1P_2P_sel} we illustrate the procedure to select the 1P and 2P stars of the outer-halo GC Arp\,2. In the left panel of the figure, we plot the $\Delta {C_{\rm F275W,F336W,F438W}}$ versus $\Delta_{\rm F275W,F814W}$ ChM of the RGB members of the cluster (gray points). 

The orange points on the bottom-left corner display the photometric errors of the ChM. Their elliptical distribution  spans few tenths of magnitudes in both the horizontal and vertical direction, thus indicating that the spread across the ChM is intrinsic. The choice to artificially plot the error distribution on the bottom left side, so that its median along the y-axis corresponds to $\Delta C_{\rm F275W, F336W, F438W} = 0$, is instrumental to the illustration of the procedure for selecting 1P stars and derive their relative fraction.

The study of the ChM of 58 GCs by \citet{milone17} has shown that 1P stars cluster around the origin of the ChM, whereas the 2P defines a sequence that extends towards larger absolute values of $\Delta {C_{\rm F275W,F336W,F438W}}$ and $\Delta_{\rm F275W,F814W}$ \citep{milone17}. For the sake of representation, we overplot a dot-dashed line on the ChM with a inclination of $18^{\circ}$ to negative direction of the x-axis. This line represents the typical inclination of the bulk of 1P stars in the ChM \citep[][see their Fig.~2]{milone17} and it has been used to fix the rotation angle of the ChM in the middle panel. Here, both the ChM and the error distribution have been rotated counter-clockwise by $18^{\circ}$, so that the dot-dashed line is now parallel to x-axis. The rotated system, with abscissa and ordinate $\Delta'_{\rm F275W,F814W}$ and $\Delta'{C_{\rm F275W,F336W,F438W}}$, respectively, allows us to identify the candidate 1P and 2P cluster stars and estimate the fraction of 1P stars with respect to the total number of RGB stars.

To do this, we plot in the right panel the histogram distributions of the $\Delta' C_{\rm F275W,F336W,F438W}$ values (gray-filled histogram) and of the errors (orange histogram). We see that the ChM histogram shows a bimodal distribution, with the bottom one associated to the 1P stars, and the top one to 2P stars. This evidence demonstrates that Arp\,2 is not chemically homogeneous.

Therefore, we calculated the $95^{\rm th}$ percentile of the error distribution along the y-axis (dotted line in the middle panel). This line delimits the portion of the ChM histogram that we have used to least-square fit a first-guess Gaussian to the 1P histogram distribution. Then we performed a new Gaussian fit but limited to the portion of the histogram with $\Delta'{C_{\rm F275W,F336W,F438W}}$ smaller than twice the dispersion of the first-guess Gaussian. Finally, we used the red-dashed line, corresponding to the peak of the best-fit Gaussian function (red curve in the right panel) shifted by twice the Gaussian dispersion, to identify the groups of bona-fine 1P and 2P stars. Finally, the ratio between the area of the best-fit Gaussian and the total area of the ChM histogram has been taken as the fraction of 1P cluster stars. In the case of Arp\,2 we find ${\rm N_{1P}/N_{TOT}} = 0.533\pm0.052$.

We applied the same procedure to build the ChM, select the candidate 1P and 2P stars and determine the corresponding ${\rm N_{1P}/N_{TOT}}$, in the other three target clusters.  

\begin{figure*}
\centering
\includegraphics[clip,trim=0.7cm 5cm 0.1cm 3cm,width=0.8\textwidth]{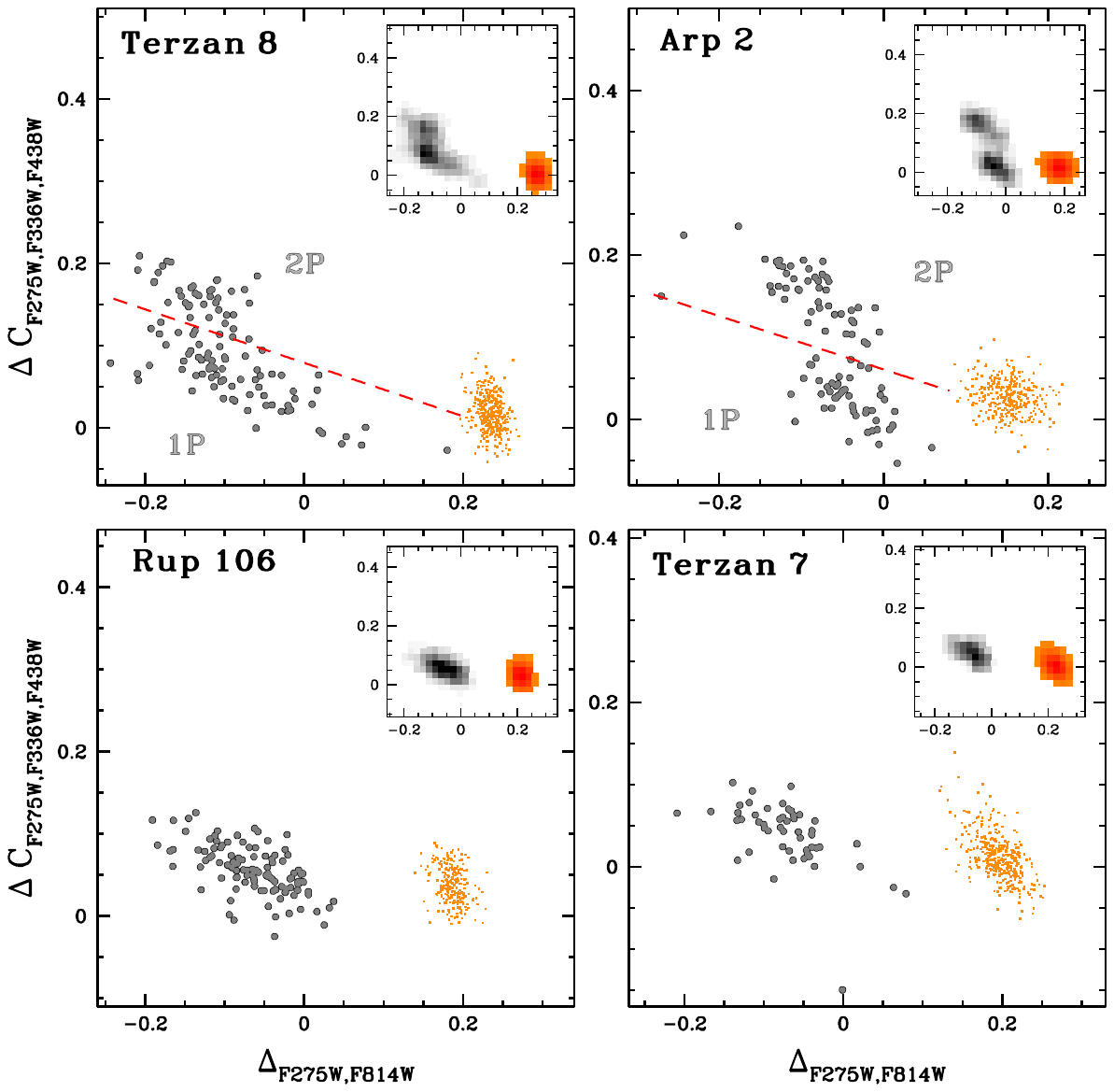}
\caption{Chromosome maps of the RGB stars in the four target clusters, arranged as in Fig.~\ref{fig:cmds_1}. In each panel, the orange points display the corresponding photometric error distribution, while the inset show the density diagram of the ChM and error distribution. The dashed line in the ChM of Terzan\,8 and Arp\,2 delimit the region of the ChM where 1P and 2P candidate stars lie. \label{fig:chms}}
\end{figure*}

The ChMs of the four target GCs are shown in Figure~\ref{fig:chms}, where we display on the bottom-right corner the corresponding photometric error distributions (orange dots). We note that in the case of Terzan\,8, Arp\,2, and Ruprecht\,106, the horizontal and vertical extension of their ChM is much larger than that of the photometric error distribution. However, in  Terzan\,7, both the horizontal and vertical spread of the ChM is comparable to those of the photometric error. 
These impressions are also confirmed by the density diagrams displayed in the inset of each panel, where the gray-shaded areas represent the density of the observed ChM points, and the orange-shaded area refers to the error distribution. 

Similarly to Fig.~\ref{fig:1P_2P_sel}, the red dashed line in the ChM of Terzan\,8 and Arp\,2 separates bona-fide 1P and 2P stars. 
As far as Ruprecht\,106 and Terzan\,7 are concerned, our procedure results in a homogeneous composition of 1P stars only. Both Terzan\,8 and Arp\,2 show a relatively high fraction of 1P stars compared to most of the studied Galactic GCs. In particular, in Terzan\,8 $\sim 71\%$ of the cluster RGB stars belong to the 1P, thus making it one of the Galactic GCs with the largest fractions of 1P stars. The 1P predominance in this cluster, identified as a 1P-dominated globulars by \citet{caloi11} based on the analysis of its HB morphology, aligns partially with the spectroscopic findings of \citet{carretta14}. \citeauthor{carretta14} estimated the 2P fraction in Terzan\,8 to be approximately 6\%, with only one out of the 16 analyzed stars showing Na abundance consistent with 2P stars. The discrepancy between this study's estimated 2P fraction (${\rm N_{2P}/N_{TOT}} \sim 29\%$) and the value reported by \citet{carretta14} likely arises from the smaller sample size used in the latter analysis. Finally, in the case of Arp\,2, 1P stars make up a fraction slightly larger than 50\% of the RGB cluster stars. 

Given the moderate population of 53 RGB stars in Terzan\,7, we conducted Monte Carlo tests to estimate the probability of detecting zero 2P stars if their fraction ranges from 1\% to 10\%. We found that if 2P stars make up 2\% of the total number of stars, the probability of detecting zero 2P stars in a sample of 53 stars is approximately 68\%. Similarly, for Ruprecht\,106, which has 109 RGB stars, the fraction of 2P stars that results in a 68\% probability of zero detections is about 1\%. Therefore, we considered 0.02 and 0.01 as our estimate of the uncertainty associated with the fraction of 2P stars in Terzan\,7 and Ruprecht\,106.

The fractions of 1P stars, the total number of RGB stars in the ChM of each target cluster, as well as their maximum radial distance as a function of the corresponding cluster half-light radius, have been reported in Table~\ref{tab:values}.

\begin{figure*}
\centering
\includegraphics[clip,trim=0.7cm 14.5cm 0.1cm 4cm,width=1\textwidth]{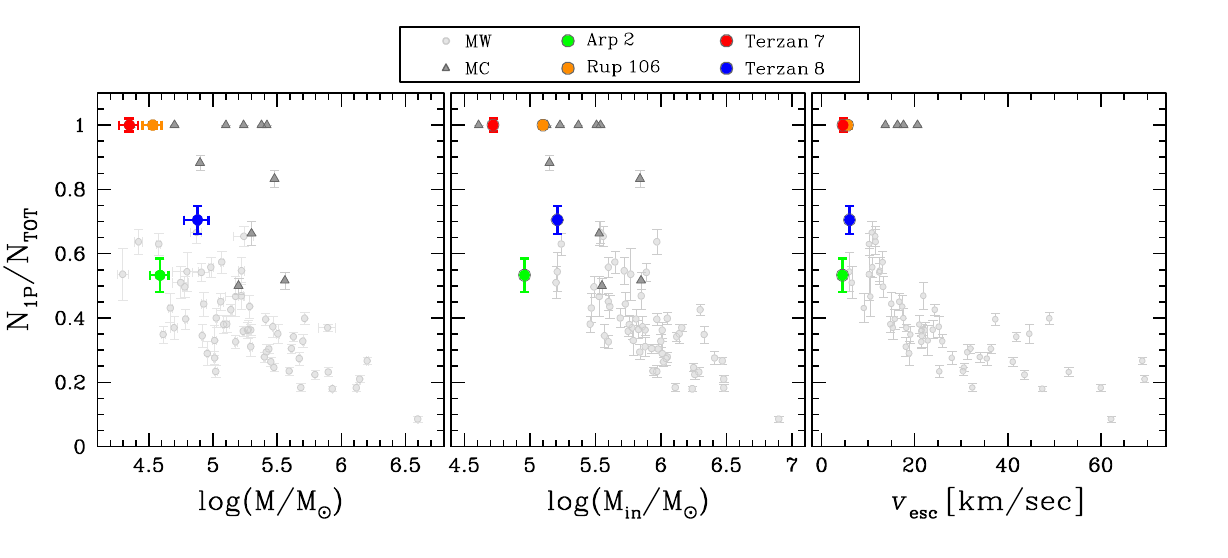}
\caption{
Fraction of 1P stars as a function of the logarithm of the GC present-day mass, initial mass, and escape velocity. The gray symbols refer to literature determinations of the 1P fractions in Milky Way and Magellanic Cloud clusters \citep{milone17,dondoglio21}, whereas the GCs studied in this paper are indicated with colored dots, as indicated in the legend. \label{fig:f1P}}
\end{figure*}

Figure~\ref{fig:f1P} shows the fraction of 1P stars as a function of the logarithm of the present-day and initial GC mass (left and middle panel, respectively) and the GC escape velocity (right panel). In addition to the GCs studied in this paper, the GC sample comprises the Galactic (gray circles) and Magellanic Cloud star clusters (gray triangles) homogeneously analyzed by using the ChMs \citep{milone17,milone20,dondoglio21}.

Terzan\,8 and Arp\,2 follow the same trend as the other Galactic GCs, thus corroborating the evidence of the anti-correlation between the fraction of 1P stars cluster and both cluster mass and escape velocity. 
The multiple-population GC with the smallest initial mass, Arp\,2 has ${\rm M_{in}} \lesssim 10^{5}$\,\msun, which is slightly lower than the empirical mass threshold of $\sim 1.5 \times 10^{5}$\,\msun suggested by \citet{milone20} as the limit for the occurrence of MPs. In the same work, Milone and collaborators have shown that the Large and Small Magellanic Cloud also host simple-population GCs with initial masses larger than the proposed threshold.

Our results represent therefore, the first photometric evidence based on the distribution of RGB stars in the ChM of the existence of SP clusters, namely Ruprecht\,106 and Terzan\,7. At the same time we provide the first photometric evidence that MPs are present in the GCs Terzan\,8 and Arp\,2.

\section{Metallicity variations among first-population stars}\label{sec:1P}

A visual inspection of the ChMs of Figure~\ref{fig:chms} reveals that the 1P stars of all clusters span a wider $\Delta_{\rm F275W,F814W}$ range than that expected from the observational uncertainties alone.

In particular, we note that, although Ruprecht\,106 does not show any vertical spread, so no presence of 2P stars, the horizontal extension of its ChM covers a $\Delta_{\rm F275W,F814W}$ interval of $\sim 0.25$\,mag, significantly wider than that of the photometric errors. Such a phenomenon is even more pronounced in the case of Terzan\,8, where the extension of the 1P candidate stars is $\sim 0.3$\,mag, while in the case of Arp\,2 1P and 2P stars seem to span a comparable interval along the x-axis. 

Recent works have shown that the horizontal width of 1P stars in the ChMs is a widespread phenomenon among Galactic GCs and demonstrated that 1P stars are not chemically homogeneous \citep{milone15a,dantona16,milone18}. Papers based on both spectroscopy and photometry have shown that the $\Delta_{\rm F275W,F814W}$ range of 1P stars is mostly due to metallicity variations \citep[e.g.][]{marino19,legnardi22,marino23}. It is important to emphasize that, while additional spectroscopic analyses are needed to confirm the presence of a metallicity spread in every GC with extended 1P sequence, alternative explanations have been proposed to account for the observed elongation of 1P stars. These alternatives include variations in helium content and the presence of unresolved interacting binaries among evolved stars, such as blue stragglers and red giants. However, the observational properties of these stars in the CMD cannot fully account for the spread observed among 1P stars in the ChM \citep{marino19a,marino23,martins20a,kamann20b,lardo22,legnardi22}.

To quantify the metallicity variations among the 1P stars of the studied clusters, we followed the method by \citet{legnardi22} and \citet{legnardi24}. Briefly, we first determined the $\Delta_{\rm F275W,F814W}$  extension of the 1G sequence as the difference between the 90$^{\rm th}$ and the 10$^{\rm th}$ percentile of the $\Delta_{\rm F275W,F814W}$ distribution. The intrinsic width, $W^{\rm 1G}_{\rm F275W,F814W}$, was estimated by subtracting the color errors in quadrature. Assuming that the observed color spread on the ChM solely derives from metallicity variations, we used the relation between the $m_{\rm F275W}-m_{\rm F814W}$ color and metallicity by \cite{dotter08a} to convert the $W^{\rm 1G}_{\rm F275W,F814W}$ values into [Fe/H] variations. Among the investigated targets, Terzan\,8 exhibits the highest [Fe/H] spread ($\delta$[Fe/H]$_{\rm 1G}^{\rm MAX}=0.373\pm0.060$), while Terzan\,7 displays the lowest value ($\delta$[Fe/H]$_{\rm 1G}^{\rm MAX}=0.036\pm0.010$). Arp\,2 and Ruprecht\,106 show moderate [Fe/H] variations, with $\delta$[Fe/H]$_{\rm 1G}^{\rm MAX}=0.087\pm0.014$ and $0.120\pm0.011$, respectively). 

\begin{figure*}
\centering
\includegraphics[width=0.8\textwidth]{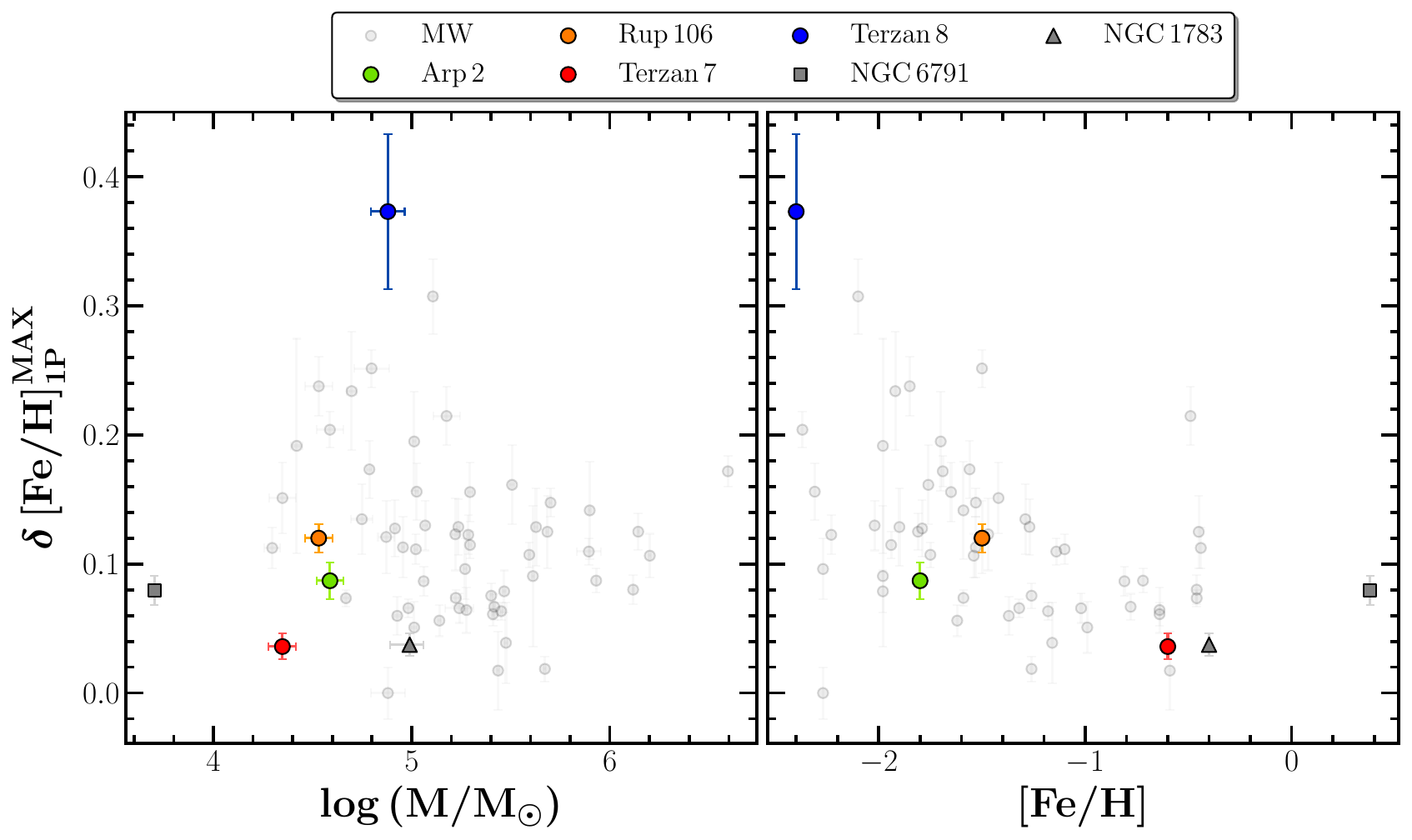}
\caption{ Maximum iron-abundance variation among 1P stars as a function of the logarithm of present-day GC mass (left) and iron abundance (right). The clusters studied in this paper are plotted with colored points as reported in the legend. The gray points indicate the GCs studied by \citet{legnardi22}, whereas the simple-population star clusters \ngc6791 and \ngc1783 are represented with a square and a triangle, respectively \citep{legnardi24}. \label{fig:1P}}
\end{figure*}

In the left panel of Figure~\ref{fig:1P} we present the maximum iron variation among 1G stars plotted against the logarithm of the cluster mass \citep[][2023 update]{baumgardt18a} and iron abundance  
\citep[][2010 update]{harris96a}. The gray dots represent the 55 Galactic GCs investigated by \citet{legnardi22}, whereas the colored symbols indicate the four targets examined in this study. The SP clusters studied by \citet{legnardi24}, namely \ngc6791 and \ngc1783, are represented with a square and a triangle, respectively. 

Our results corroborate the evidence of a mild correlation between $\delta$[Fe/H]$^{\rm 1G}_{\rm MAX}$ and the metallicity, and between $\delta$[Fe/H]$^{\rm 1G}_{\rm MAX}$ and mass. Indeed, the $\delta$[Fe/H]$_{\rm 1G}^{\rm MAX}$ values obtained for Arp\,2, Ruprecht\,106, and Terzan\,7 follow the general trend in both the diagrams. Conversely, 1G stars in Terzan\,8 exhibit higher metallicity variations compared to Galactic GCs with similar mass/metallicity.

Moreover, the evidence of extended 1P sequences in the ChMs of Ruprecht\,106 and Terzan\,7 shows that the metallicity variations are not peculiar to GCs with MPs but also characterize SP clusters.

\begin{deluxetable*}{cccccc}
\tabletypesize{\small} 
\tablecaption{Values of the RGB widths, fraction of 1P candidate stars, total number of stars analyzed in the chromosome map, and maximum radial distance from the cluster center ($r_{\rm max}$) as a fraction of the cluster half-light radius ($r_{\rm hl}$). \label{tab:values}}
\tablehead{\colhead{Cluster} & \colhead{$W_{\rm F275W,F814W}$} & \colhead{$W_{\rm C_{F275W,F336W,F438W}}$} & \colhead{$N_{\rm 1P}/N_{\rm TOT}$} & \colhead{$N_{\rm TOT}$} & \colhead{$r_{\rm max}/r_{\rm hl}$} \\
  & (mag) & (mag) &  &  & }
\startdata
Arp\,2        & $0.111\pm0.096$ & $0.196\pm0.024$ & $0.533\pm0.052$ &  96 & 0.98 \\
Ruprecht\,106 & $0.125\pm0.006$ & $0.100\pm0.020$ & $1.000\pm0.010$ & 109 & 1.25 \\ 
Terzan\,7     & $0.152\pm0.051$ & $0.081\pm0.015$ & $1.000\pm0.020$ &  53 & 1.66 \\
Terzan\,8     & $0.203\pm0.021$ & $0.188\pm0.024$ & $0.705\pm0.044$ & 114 & 0.84 \\
\enddata
\end{deluxetable*}

\section{Summary and conclusions}\label{sec:summary}
We used multi-epoch ACS/WFC and UVIS/WFC3 images in F275W, F336W, F438W, F606W, and F814W from the GO\,17075 (PI\,Lagioia) and from the \hst\ archive to investigate MPs in the four outer-halo GCs, Arp\,2, Ruprecht\,106, Terzan\,7, and Terzan\,8. We derived high-precision stellar magnitudes, positions, and PMs by using the methods and the computer programs developed by Jay Anderson \citep[e.g.][]{anderson00, anderson06}. The resulting five-band photometry is comparable to that of 58 less distant GCs studied in the recent surveys of MPs \citep{piotto15,milone17}, thus providing a total sample of 62 Galactic GCs homogeneously analyzed. The main results are summarized in the following.

\begin{enumerate}

\item We derived the $\Delta_{C \rm F275W,F336W,F438W}$ versus $\Delta_{\rm F275W,F814W}$ ChMs to search for MPs with different light-element abundances. We find that Arp\,2 and Terzan\,8 host MPs, whereas Ruprecht\,106 and Terzan\,7 are SP clusters. This is the first evidence of simple stellar population GCs based on the ChM, and corroborates previous conclusions based on spectroscopy of nine stars of Ruprecht\,106 \citep{villanova13}, photometry in the U, B, and I bands of RGB stars in both Ruprecht\,106 and Terzan\,7 \citep{dotter18,lagioia19a}, and the HB morphology of Ruprecht\,106 \citep{milone14}.

\item It is widely accepted that the $C_{\rm F275W, F336W, F438W}$ pseudo-color width of the RGB correlates with the average cluster metallicity. In GCs with MPs, the RGB width mostly depends on the maximum internal variation in the abundances of CNO. This color index is primarily sensitive to the nitrogen content of the stellar populations and, to a lesser extent, to variations in helium, carbon, and oxygen \citep[e.g.][]{milone15a,lagioia19a,jang21}. In addition to light-element spread, star-to-star metallicity variations can also contribute to the RGB width \citep[e.g.][]{marino19a}.
We find that the observed RGB widths of Arp\,2 and Terzan\,8 follow the same trend as the bulk of Galactic GCs in the $W_{C \rm F275W,F336W,F438W}$ versus [Fe/H] diagram. On the contrary, the RGB width of Ruprecht\,106 and Terzan\,7 are much narrower than those of GCs with similar [Fe/H] as expected for clusters with homogeneous abundance of light-elements. 

\item Previous works have suggested that a mass threshold could govern the occurrence of MPs in GCs. Based on analysis of the ChMs in 58 Galactic GCs, \citet{milone20} noticed that the MP GCs have masses larger than $\sim 1.5\times10^5$\,\msun. Arp\,2 is the Galactic GC with the smallest initial mass ($\sim 0.9 \times 10^{4}$\,\msun) for which MPs have been detected through the ChMs. The discovery of MPs in this cluster suggests that the proposed mass threshold, if present, should be smaller than $\sim 10^5$\,\msun. In this context, it is worth noticing that various SP Magellanic Cloud GCs have initial masses of 1 -- 3 $\times 10^5$\,\msun, which are larger than those of Arp\,2 and Terzan\,8. This fact could indicate that, in addition to cluster mass, the GC formation environment, affects the occurrence of MPs with metal-poor GCs having a lower initial-mass threshold for the formation of 2P stars \citep{krause16,huang24}. In this regard, a possible interpretation for the non occurrence of MPs in some GCs is that the emergence of star clusters with MPs may also depend on the environment \citep{krause16,huang24}, though it is not obvious how the environment could affect the process. An alternative explanation for the lack of MPs is that this phenomenon would only occur in stars less massive than $\sim 1.6$\,\msun only \citep{bastian18}. In such a case case, the reason for the lack of detection of 2P stars in some Magellanic Cloud GCs is that MP studies in Magellanic Cloud GCs younger than $\sim 2$\,Gyr are exclusively based on RGB stars. 

\item  We used the ChM to derive the fractions of 1P stars in Arp\,2 and Terzan\,8, which correspond to 53.3$\pm$5.2\% and 71.0$\pm$4.4\%, respectively. Noticeably, the Galactic GCs with MPs and large perigalactic radii (${\rm R_{peri}} > 3.5$\,kpc) studied in the literature tend to have higher fractions of 1P stars compared to the other Milky Way GCs. This behavior is particularly pronounced in the clusters studied in the MC, where the fraction of 1P stars is even higher than in Galactic GCs with large ${\rm R_{peri}}$ \citep{zennaro19, milone20}, thus suggesting that the fraction of 1P stars may depend on the environment. Intriguingly, while Terzan\,8 clearly follows the overall trend of MW GCs, Arp\,2 is a possible outlier and shares similar fractions of 1P stars as some Galactic GCs with similar masses and small ${\rm R_{peri}}$.

\item The 1P stars of the multiple-population GCs Arp\,2 and Terzan\,8 exhibit intrinsic $m_{\rm F275W}-m_{\rm F814W}$ color spreads, which likely indicates the presence of star-to-star [Fe/H] variations of $\sim 0.1$ and 0.3\,dex, respectively.
A recent study of 55 Galactic GCs with MPs by \citet{legnardi22} has shown that the 1P stars of virtually all clusters exhibit extended sequences in the ChM. These observations are consistent with inhomogeneous [Fe/H] abundance, ranging from less than 0.05\,dex to $\sim 0.3$\,dex. Hence, the results on Arp\,2 and Terzan\,8 corroborate the evidence that metallicity variations among 1P stars are common features of GCs. 
    
\item We found that the SP GCs Ruprecht\,106 and Terzan\,7 also exhibit extended 1P sequences in the ChM, closely resembling those observed in 1P stars of MP GCs. This finding, combined with evidence of star-to-star [Fe/H] variations in the open cluster \ngc6791 and the $\sim 1.5$ Gyr-old star cluster \ngc1783 \citep{legnardi24}, indicates that this phenomenon is not exclusive to MP GCs but also occur in SP GCs, open clusters, and intermediate-age massive Magellanic Cloud clusters. Although high-precision spectroscopic observations are required to definitively determine the chemical composition of these systems, star-to-star metallicity variations remain the most plausible explanation for the extended sequences observed in their ChMs.

\end{enumerate}

\begin{acknowledgments}
E. P. Lagioia acknowledges support from the "Science \& Technology Champion Project" (202005AB160002) and from the "Top Team Project" (202305AT350002), all funded by the "Yunnan Revitalization Talent Support Program".
T. Ziliotto acknowledges funding from the European Union’s Horizon 2020 research and innovation program under the Marie Skłodowska-Curie Grant Agreement No. 101034319 and from the European Union – Next Generation EU.
This work has received funding from "PRIN 2022 2022MMEB9W - {\it Understanding the formation of globular clusters with their multiple stellar generations}" (PI Anna F.\,Marino), and from INAF Research GTO-Grant Normal RSN2-1.05.12.05.10 -  (ref. Anna F. Marino) of the "Bando INAF per il Finanziamento della Ricerca Fondamentale 2022".
\end{acknowledgments}

%

\vspace{5mm}
\facilities{HST(WFC3/UVIS), HST(WFC/ACS)}

%


\software{
          }

\bibliography{ms.bib}{}
\bibliographystyle{aasjournal}



\end{document}